\input harvmac


%
%

\font\zfont = cmss10 
 
\def\bigone{\hbox{1\kern -.23em {\rm l}}} \def\ZZ{\hbox{\zfont Z\kern-.4emZ}}         

\def\a{\alpha}
\def\b{\beta}
\def\g{\gamma}
\def\d{\delta}
\def\e{\epsilon}
\def\et{\eta}
\def\th{\theta}

\def\p{\phi}

\def\c{\chi}

\def\O{\Omega}
\def\o{\over}

\def\ea{{\eta_1}}
\def\eb{{\eta_2}}
\def\eab{{\bar{\eta_1}}}
\def\ebb{{\bar{\eta_2}}}
\def\vb{{\bar{v}}}

\def\bal{{\bar \alpha}}

\def\w{\wedge}

\def\bz{{\bar z}}
\font\cmss=cmss10 \font\cmsss=cmss10 at 7pt
\def\rlx{\relax\leavevmode}

\def\RR{\relax{\rm I\kern-.18em R}}
\def\ZZ{\rlx\leavevmode\ifmmode\mathchoice{\hbox{\cmss Z\kern-.4em
Z}}{\hbox{\cmss Z\kern-.4em Z}}{\lower.9pt\hbox{\cmsss Z\kern-.36em Z}}
{\lower1.2pt\hbox{\cmsss Z\kern-.36em Z}}\else{\cmss Z\kern-.4em Z}\fi}

\def\ap{\alpha'}

\def\frac#1#2{{#1 \over #2}}
\def\pa{\partial}
\def\dg{\dagger}
\def\rd{{ d}}

\def\bpa{{\bar \partial}}

\def\bb{{\bar b}}
\def\bc{{\bar c}}

\def\bC{{\bar C}}
\def\bD{{\bar D}}

\def\dza{dz_1}
\def\dzb{dz_2}
\def\dzc{dz_3}
\def\dzd{dz_4}
\def\za{z_1}
\def\zb{z_2}

\def\dbza{d{\bar z_1}}
\def\dbzb{d{\bar z_2}}
\def\dbzc{d{\bar z_3}}
\def\dbzd{d{\bar z_4}}

\def\om{{\omega}}
\def\bom{{\bar \omega}}

\def\na{\nabla}

\def\IZ{\relax\ifmmode\mathchoice
{\hbox{\cmss Z\kern-.4em Z}}{\hbox{\cmss Z\kern-.4em Z}}
{\lower.9pt\hbox{\cmsss Z\kern-.4em Z}} {\lower1.2pt\hbox{\cmsss
Z\kern-.4em Z}}\else{\cmss Z\kern-.4em Z}\fi}


\lref\AGM{P.~S.~Aspinwall, B.~R.~Greene and D.~R.~Morrison,
``Multiple mirror manifolds and topology change in string theory,''
Phys.\ Lett.\  B {\bf 303}, 249 (1993) [arXiv:hep-th/9301043].
} \lref\WitP{E.~Witten, ``Phases of N = 2 theories in two
dimensions,'' Nucl.\ Phys.\  B {\bf 403}, 159 (1993)
[arXiv:hep-th/9301042].
}

\lref\StromC{A.~Strominger, ``Massless black holes and conifolds in
string theory,'' Nucl.\ Phys.\  B {\bf 451}, 96 (1995)
[arXiv:hep-th/9504090].
}

\lref\GMS{B.~R.~Greene, D.~R.~Morrison and A.~Strominger, ``Black
hole condensation and the unification of string vacua,'' Nucl.\
Phys.\  B {\bf 451}, 109 (1995) [arXiv:hep-th/9504145].
}

\lref\Strom{ A.~Strominger, ``Superstrings with torsion,''
Nucl.\ Phys.\ B {\bf 274}, 253 (1986). }
\lref\Bars{ I.~Bars,
``Compactification of superstrings and torsion,''
Phys.\ Rev.\ D {\bf 33}, 383 (1986); 
I.~Bars, D.~Nemeschansky and S.~Yankielowicz,
``Compactified superstrings and torsion,''
Nucl.\ Phys.\ B {\bf 278}, 632 (1986).
}
\lref\Hull{ C.~M.~Hull, ``Anomalies, ambiguities and superstrings,''
Phys.\ Lett.\ B {\bf 167}, 51 (1986).
}
\lref\Hulla{ C.~M.~Hull,
``Sigma model beta functions and string compactifications,''
Nucl.\ Phys.\ B {\bf 267}, 266 (1986);
C.~M.~Hull and E.~Witten,
``Supersymmetric sigma models and the heterotic string,''
Phys.\ Lett.\ B {\bf 160}, 398 (1985);
}

\lref\FY{J.-X.~Fu and S.-T.~Yau,
``The theory of superstring with flux on non-K\"ahler manifolds and the
complex Monge-Amp\'ere equation,''
[arXiv:hep-th/0604063].
}

\lref\BBFTY{K.~Becker, M.~Becker, J.-X.~Fu, L.-S.~Tseng and S.-T.~Yau,
``Anomaly cancellation and smooth non-K\"ahler solutions in heterotic
string theory,''Nucl.\ Phys.\ B {\bf 751}, 108 (2006)
[arXiv:hep-th/0604137].
}

\lref\GMW{ J.~P.~Gauntlett, D.~Martelli and D.~Waldram,
``Superstrings with intrinsic torsion,''
Phys.\ Rev.\  D {\bf 69}, 086002 (2004)
[arXiv:hep-th/0302158].
}

\lref\Dall{ G.~Dall'Agata, ``On supersymmetric solutions of type
IIB supergravity with general fluxes,''
Nucl.\ Phys.\  B {\bf 695}, 243 (2004)
[arXiv:hep-th/0403220].
}

\lref\GKMW{J.~P.~Gauntlett, N.~Kim, D.~Martelli and D.~Waldram,
``Fivebranes wrapped on SLAG three-cycles and related geometry,''
 JHEP {\bf 0111}, 018 (2001)
 [arXiv:hep-th/0110034].
}

\lref\dbeck{K.~Dasgupta, G.~Rajesh and S.~Sethi,
``M theory, orientifolds and G-flux,''  JHEP {\bf 9908}, 023 (1999)
[arXiv:hep-th/9908088].
   }
\lref\BD{K.~Becker and K.~Dasgupta, ``Heterotic strings with torsion,''
   JHEP {\bf 0211}, 006 (2002)
   [arXiv:hep-th/0209077].
}

\lref\beck{K.~Becker, M.~Becker, K.~Dasgupta and P.~S.~Green,
 ``Compactifications of heterotic theory on non-K\"ahler complex manifolds I,''
 JHEP {\bf 0304}, 007 (2003), [arXiv:hep-th/0301161];
 K.~Becker, M.~Becker, P.~S.~Green, K.~Dasgupta and E.~Sharpe,
 ``Compactifications of heterotic strings on non-K\"ahler complex manifolds II,''
  Nucl.\ Phys.\ B {\bf 678}, 19 (2004), [arXiv:hep-th/0310058].}

\lref\BT{K.~Becker and L.-S.~Tseng, ``Heterotic flux
compactifications and their moduli,'' Nucl.\ Phys.\  B {\bf 741},
162 (2006) [arXiv:hep-th/0509131].
}

\lref\BTY{ M.~Becker, L.~S.~Tseng and S.~T.~Yau, ``Moduli space of
torsional manifolds,'' [arXiv:hep-th/0612290].
}

\lref\BBDTA{K.~Becker, M.~Becker, K.~Dasgupta and R.~Tatar,
``Geometric transitions, non-K\"ahler geometries and string vacua,''
Int.\ J.\ Mod.\ Phys.\  A {\bf 20}, 3442 (2005)
[arXiv:hep-th/0411039];
M.~Becker, K.~Dasgupta, S.~H.~Katz, A.~Knauf and R.~Tatar,
``Geometric transitions, flops and non-K\"ahler manifolds. II,''
Nucl.\ Phys.\  B {\bf 738}, 124 (2006)
[arXiv:hep-th/0511099].
}

\lref\Adams{A.~Adams, ``Conformal field theory and the Reid
conjecture,'' [arXiv:hep-th/0703048].
}

\lref\AEL{ A.~Adams, M.~Ernebjerg and J.~M.~Lapan, ``Linear models
for flux vacua,'' [arXiv:hep-th/0611084].
}

\lref\Vafa{C.~Vafa, ``Superstrings and topological strings at large
N,'' J.\ Math.\ Phys.\  {\bf 42}, 2798 (2001)
[arXiv:hep-th/0008142].}

\lref\gop{R.~Gopakumar and C.~Vafa, ``On the gauge theory/geometry
correspondence,''  Adv.\ Theor.\ Math.\ Phys.\  {\bf 3}, 1415 (1999)
[arXiv:hep-th/9811131].}

\lref\CKT{W.-Y.~Chuang, S.~Kachru and A.~Tomasiello, ``Complex/symplectic mirrors '' [arXiv:hep-th/0510042].}

\lref\ati{M.~Atiyah, J.~M.~Maldacena and C.~Vafa, ``An M-theory flop
as a large N duality,'' J.\ Math.\ Phys.\  {\bf 42}, 3209 (2001)
   [arXiv:hep-th/0011256].}

\lref\reid{M.~Reid, ``The moduli space of 3-folds with $K=0$ may
nevertheless be irreducible'', \ Math. \ Ann. {\bf 278}, 329
(1987).}

\lref\clemens{H.~Clemens, ``Double solids,'' Adv.\ Math.\ {\bf 47},
107 (1983).}

\lref\friedman{R.~Friedman, ``Simultaneous resolution of threefold
double points,'' Math.\ Ann.\ {\bf 274} 671 (1986).}

\lref\BB{K.~Becker and M.~Becker, ``M-Theory on Eight-Manifolds,''
Nucl.\ Phys.\  B {\bf 477}, 155 (1996) [arXiv:hep-th/9605053].
}

\lref\SVW{S.~Sethi, C.~Vafa and E.~Witten, ``Constraints on
low-dimensional string compactifications,'' Nucl.\ Phys.\  B {\bf
480}, 213 (1996) [arXiv:hep-th/9606122].
} \lref\Sen{A.~Sen, ``F-theory and Orientifolds,'' Nucl.\ Phys.\  B
{\bf 475}, 562 (1996) [arXiv:hep-th/9605150].
}

\lref\DVV{R.~Dijkgraaf, E.~P.~Verlinde and H.~L.~Verlinde, ``C = 1
Conformal Field Theories on Riemann Surfaces,'' Commun.\ Math.\
Phys.\  {\bf 115}, 649 (1988).
}

\lref\Ginsp{P.~H.~Ginsparg, ``Curiosities at c = 1,'' Nucl.\ Phys.\
B {\bf 295}, 153 (1988).
}

\lref\AspinK{P.~S.~Aspinwall, ``K3 surfaces and string duality,''
[arXiv:hep-th/9611137].
}

\lref\AspinD{P.~S.~Aspinwall, ``An analysis of fluxes by duality,''
[arXiv:hep-th/0504036].
}

\lref\AspinP{P.~S.~Aspinwall,
``Point-like instantons and the spin(32)/Z(2) heterotic string,''
Nucl.\ Phys.\  B {\bf 496}, 149 (1997)
[arXiv:hep-th/9612108].
}

\lref\SY{S.~Kim and P.~Yi,
``A heterotic flux background and calibrated five-branes,''
JHEP {\bf 0611}, 040 (2006)
[arXiv:hep-th/0607091].
}



\Title{ {\vbox{
}}}
{\vbox{ \hbox{
\centerline{Heterotic K\"ahler/non-K\"ahler Transitions}}\hbox{}
\hbox{\centerline{}} }}
\bigskip
\bigskip
\centerline{Melanie Becker$^{1,2}$, ~Li-Sheng Tseng$^{2,3}$~
and~Shing-Tung Yau$^3$}
\bigskip
\bigskip
\centerline{$^1$ \it George P. and Cynthia W. Mitchell Institute for
Fundamental Physics}
\centerline{\it Texas A \& M University, College Station, TX 77843, USA}
\smallskip
\centerline{$^2$\it
Jefferson Physical Laboratory, Harvard University, Cambridge, MA
02138, USA}
\smallskip
\centerline{$^3$\it Department of Mathematics, Harvard University,
Cambridge, MA 02138, USA}

\bigskip


\bigskip

\bigskip

\bigskip
\centerline{\bf Abstract}

\bigskip

We show how two topologically distinct spaces - the K\"ahler $K3
\times T^2$ and the non-K\"ahler $T^2$ bundle over $K3$ - can be
smoothly connected in heterotic string theory. The transition occurs
when the base $K3$ is deformed to the $T^4/\ZZ_2$ orbifold limit. The
orbifold theory can be mapped via duality to M-theory on $K3\times
K3$ where the transition corresponds to an exchange of the two
$K3$'s.

\bigskip
\baselineskip 18pt
\bigskip
\noindent

\Date{June, 2007}


\newsec{Introduction}

Background geometry affects strings and point particles very
differently. From the many well-studied string dualities, we know
that string theories on different geometrical spaces can be dual,
that is identical up to some identification.  Moreover, string
theory can also smoothly connect topologically different spaces.
Some of these string transitions, for example those of flops
\refs{\AGM,\WitP,\ati} and conifolds \refs{\StromC,\GMS,\gop,
\Vafa}, have geometrical origins and are closely related to the
mathematics of singularity resolutions.

It is conceivable that many vacua in the string theory landscape are
connected by transitions and one may wonder if the more recently
studied manifolds with torsion can be connected to more conventional
string theory compactifications. Many years ago, it was conjectured
\reid\ that Calabi-Yau manifolds can indeed be connected via
transitions through non-K\"ahler manifolds \refs{\clemens,\friedman}
(see also \CKT).  K\"ahler/non-K\"ahler transitions have also been
described recently from the worldsheet conformal field theory point of view
in \Adams\ in the context of gauged linear sigma models \AEL.

In this paper, we take the space-time approach to explore transitions
between K\"ahler and non-K\"ahler manifolds in the context of flux
compactifications of heterotic string theory.\foot{For non-compact
geometries, non-K\"ahler to non-K\"ahler
transitions in heterotic strings have been discussed in \BBDTA}
In the heterotic theory,
there can be two types of fluxes - the gauge two-form $F_{MN}$ and
the three-form $H_{MNP}$. Preserving supersymmetry in four
dimensions will constrain both the compactification geometry and
the fluxes.  It is the goal of this paper to relate two different
types of spaces both of which are locally $K3 \times T^2$.  The
first is the geometry of
the $K3\times T^2$ manifold with non-zero $U(1)$ gauge fields.  The
second is the non-K\"ahler geometry of a $T^2$ bundle over $K3$, the
FSY (Fu-Strominger-Yau) geometry.

This paper is organized as follows. In Section~2, we analyze the
conditions under which a Kummer surface can be blown down to a
$T^4/\ZZ_2$ orbifold in the presence of fluxes while maintaining
supersymmetry throughout. In Section~3, we show that a transition
between the K\"ahler geometry $K3 \times T^2$ and the non-K\"ahler
FSY geometry can take place using the mapping to M-theory on $K3\times
K3$ where the transition corresponds to an exchange of the two $K3$'s.
In Section~4, we present our conclusions.  In an appendix, we work out
the conditions for the FSY geometry to preserve $N=2$ supersymmetry, which
is necessary in order that a smooth transition to the $K3\times T^2$
geometry can take place.

\newsec{$N=2$ heterotic compactifications}

We are interested in supersymmetric compactifications to four
dimensions in heterotic string theory.  The conditions on the
hermitian form $J$ and the holomorphic (3,0)-form $\O$ are \foot{We mostly
follow the notation and conventions of \BBFTY.  In this paper, $F=F^a T^a$ is taken to be anti-hermitian with ${\rm tr}\,T^aT^b = -\d^{ab}$.  For convenience, we
shall subsequently also set $2\pi\sqrt{\a'}=1$.}
\eqn\balance{\rd(\|\Omega\|_J\, J\w J)=0~,}
\eqn\hermitym{F^{(2,0)}=F^{(0,2)}=0~,\quad F_{mn}J^{mn}=0~,}
\eqn\anomcan{2i\, \pa\bpa J = {\ap \o 4} [{\rm tr} (R\w R) - {\rm
tr}(F\w F)]~.}
The standard background fields - the three-form $H$
and the dilaton $\p$ - are determined from the supersymmetric
constraints \eqn\Hdef{H=i(\bpa-\pa) J~,}
\eqn\Omdef{\|\Omega\|_J=e^{-2(\p+\p_0)}~.}

We focus on two special classes of solutions.  The first is the $K3
\times T^2$ solution \Strom.  The hermitian metric and the
holomorphic three-form is taken to be
\eqn\KTsol{J = e^{2\phi} J_{K3} + \frac{i}{2}\, dz \w d\bz~,
\qquad\qquad~~\qquad\qquad \O = \O_{K3}\w dz~.}
In general, $\phi$ is non-constant and has dependence on the $K3$
coordinates. By \Hdef, this gives a contribution to the $H$-field.
The second is the torus bundle over $K3$ solution \refs{\beck,
\FY,\BBFTY} with the torus twisted with respect to the $K3$ base.
The metric and the three-form are generalized to
\eqn\FYsol{J = e^{2\phi} J_{K3} + \frac{i}{2} (dz+\a)\w
(d\bz + {\bar \a})~, \qquad\qquad \O=\O_{K3}\w \th~,}
where $\th=(dz+\a)$ is a globally-defined $(1,0)$-form.
We shall only consider the case where the curvature of the torus
bundle $\om=\om_1+i\om_2=d \th = d \a \in H^2(K3,\ZZ)$ is
a $(1,1)$-form; that is,
\eqn\omdef{\om^{(2,0)}=\om^{(0,2)}=\om_{mn}J_{K3}{}^{mn}=0~.}
The $H$-field now has contributions from both the derivative of
$\phi$ and also $\om$.

To fully describe both solutions, we have to specify the gauge bundle.
In addition to being hermitian-Yang-Mills \hermitym, the gauge
bundle must satisfy the anomaly equation \anomcan.  Integrating
the anomaly equation over $K3$ leads to the topological condition
\eqn\kfcon{ \frac{1}{16\pi^2}\int_{K3}{\rm tr}\, F\w F - \int_{K3}
\om\w \bom= 24 ~,}
where $\bom= \om_1-i\om_2$ is the complex conjugate of $\om$.  It
turns out that this condition is sufficient to guarantee that the
anomaly equation, which for these geometries can be interpreted
as a highly non-linear second order partial differential equation
for $\phi$, can be solved.  \refs{\FY,\BBFTY}.  Note that by
\hermitym\ and \omdef, both $F$ and
$\om$ are anti-self-dual and therefore the left hand side of \kfcon\
is positive semi-definite.  As a simplification, we will consider
only direct sums of $U(1)$ gauge bundles.  Dirac quantization
requires that $\frac{i}{2\pi}F^a\in H^{2}(K3,\ZZ)$.  Considering also
\hermitym\ and \omdef, the field strength $F$ and $\om$ indeed
satisfy identical equations.  This is suggestive that the the gauge
bundle and torus bundle under appropriate conditions might perhaps be
interchangeable.  In general, this is not the case.  But we shall show in
Section~3 that different values for the pair $(F, \om)$ can be
smoothly connected.

We point out that the $K3 \times T^2$ solution preserves $N=2$
SUSY in four dimensions. Likewise, with $\om \in H^{(1,1)}(K3,\ZZ)$,
the FSY geometry also preserves $N=2$ SUSY.  These two solutions must
preserve the same amount of supersymmetry if we desire a smooth
transition between them.  It is worthwhile to emphasize that the
conditions for spacetime supersymmetry do allow for the presence
of a $(2,0)$ component for $\om$.  However, the resulting
four-dimensional supersymmetry  would then be reduced down to $N=1$.
A discussion of the supersymmetry of the FSY geometry is provided
in Appendix A.

\subsec{Deforming to the orbifold limit of $K3$}

We will take the $K3$ surface $S$ to be a Kummer surface. This can
be described as the blow-up of $T^4/\ZZ_2$ at all 16 fixed points.
Here, we want to find the conditions under which a Kummer surface
can be blown down to a $T^4/\ZZ_2$ orbifold in the presence of fluxes
while maintaining supersymmetry throughout.  Keeping the complex
structure fixed, the supersymmetry variation conditions with
non-zero fluxes were worked out in \BTY.  For ease of presentation, we
describe below the equivalent reverse process of blowing up the
fixed points for a given flux.

Let $C_i\in H_2(S,\ZZ)$, $i=1,\ldots, 16$, be a basis for the 16
blown up (-2)-curves of the Kummer surface.\foot{For a review of
the mathematical aspects of $K3$ surfaces, see \AspinK.}  We let
$\b_i \in H^2(S,\ZZ)$ be the associated dual two-forms.  Since
these rational curves are localized and thus disjoint, the
matrix of intersection numbers is
\eqn\tnorm{C_i.C_j= \int_S \b_i \w \b_j = -2\,\d_{ij}~. }
Note that this intersection matrix is different from that for
the standard basis of $H_2(S,\ZZ)$ which is given by
\eqn\Htmet{(-E_8)\otimes (-E_8) \otimes \left(\matrix{0&1\cr 1&0}\right)
\otimes \left(\matrix{0&1\cr 1&0}\right) \otimes
\left(\matrix{0&1\cr 1&0}\right)~, }
where $E_8$ denotes the Cartan matrix of the Lie algebra $E_8$.
Though the $16$ $C_i$'s together with the 6 two-cycles of
$T^4/\ZZ_2$ provide a natural set of elements in $H_2(S, \ZZ)$
for the Kummer surface, this set as constituting a basis turns
out to only generate a sublattice of the full $H_2(S,\ZZ)$ lattice
(see for example \AspinP).  This ``Kummer" basis however can be used as a
basis for $H_2(S,{\bf Q})$.


Proceeding on, the area of the $i^{th}$-rational curve is given
by\foot{We will take $J=J_{K3}$ in this subsection since we are
only interested here in deforming the $K3$.}
\eqn\arat{A_i=\int_{C_i} J = \int_S J \w \b_i ~,}
where we have used the dual relation associating $C_i \sim \b_i$.
In the orbifold limit, each rational curve shrinks to a point and
thus $A_i=0$. To deform away from the orbifold limit, we want to
deform $J$ such that $\d A_i > 0$.  That is we need,
\eqn\adeform{\int_{C_i} J+ \d J = \int_{C_i} \d J = \int_S \d J \w
\b_i >0~.}
Furthermore, to preserve supersymmetry in the presence of fluxes,
{\it i.e.} non-zero torus bundle and/or gauge bundle curvature, we have
to satisfy the additional conditions \BTY
\eqn\susyreq{\int_S \d J \w \om =0~, \qquad \qquad \int_S
\d J \w F = 0~.}
These arise from varying the primitivity condition for the
curvatures $\om$ and $F$.  Let us focus below on the torus bundle
curvature as the conditions for the the gauge bundle are identical.
Varying $\om \w J= 0$ implies
\eqn\susyexp{\eqalign{0&=\d \om \w J + \om \w J\cr &= i\pa\bpa f
\w J + \om\w \d J~,}}
where $\d \om = i\pa\bpa f$ imposes that $\om$ can only vary in
its cohomology class.  Taking the hodge star of \susyexp\ results
in $\Delta f = *(\om \w \d J)$.  This then implies the integral condition in
\susyreq, which is the necessary and sufficient condition that a
solution for $f$ exists.

It is not difficult to satisfy both \adeform\ and \susyreq.  Using the $\b_i$ basis, let $ \d J = a^j \b_j$ and $\om = b^i \b_i\,$.\foot{Here, we have suppressed the non-relevant terms associated with the six nonlocalized (1,1)-forms on $T^4/\ZZ_2$ which are orthogonal to the localized forms $\b_i$.}  Then, we have
\eqn\brat{\int_{C_i}\d J = \int_S a^j\b_j \w \b_i = -2\, a_i = \d A_i >0~,}
\eqn\bvar{\int_S \d J \w \om = \int_S a^j \b_j \w b^i \b_i =
-2\,a_i b^i = \d A_i b^i = 0~,}
where $\d A_i = -2\,a_i$.  Thus, as long as two of the $b^i\,$'s are non-zero and also not of the same signs, there exist positive $\d A_i$'s as required for \brat\ that satisfy \bvar.  This is the condition for $\om$ (and similarly for $F$) that ensures that the singularities can be blown up.

As a simple example, let $\om$ for the FSY model be given by
\eqn\fya{\om=\om_1+i\om_2 = (1+i)\b_1 + (1+i)\b_2 - 2(1+i) \b_3~,}
and $F=0$.  With \tnorm, this model satisfies that anomaly condition \kfcon
\eqn\omanoma{-\int_S \om \w \bom = -\int_S 2(\b_1\w \b_1 + \b_2 \w \b_2)
+ 8\,\b_3\w\b_3 =24~.}
It can be easily checked that \brat\ and \bvar\ are satisfied for
$\d A_1=\d A_2=\d A_3=a>0$, where $a$ is an arbitrary positive constant.
$\d A_i>0$ for $i=4,\ldots, 16$ are not constrained.  We shall show in the next section
that this FSY model at the orbifold limit $K3=T^4/\ZZ_2$ is smoothly connected to the
$K3\times T^2$ model with non-zero $U(1)$ gauge field strengths
\eqn\kta{\frac{F^1}{2\pi}=\frac{F^2}{2\pi}= -\frac{F^3}{4\pi}=
(1+i)\,\dbza\w\dzb + (1-i)\,\dza\w\dbzb,}
where the superscript index in $F^i$, $i=1,\ldots,16$, denotes the $16~U(1)$ gauge
field strengths and $(\za,\zb)$ are the complex coordinates on $T^4/\ZZ_2$.


\newsec{Duality at the orbifold point}

At the $T^4/\ZZ_2$ orbifold of $K3$, we can map the heterotic solutions
to those of M-theory on $Y=K3 \times K3$, where each $K3$ is a $T^4/\ZZ_2$ orbifold.  Let us recall the chain of dualities \refs{\dbeck, \BD}.  Starting from M-theory compactified on
$Y$, we can treat the second  $T^4/\ZZ_2$ orbifold as a torus
fibration over $T^2/\ZZ_2$.  Taking the area of the torus fiber to zero, we arrive at
the type IIB theory on an orientifold $T^4/\ZZ_2 \times T^2/\ZZ_2$,
where at each of the four fixed points of $T^2/\ZZ_2$, there are
four D7 branes and one O7 brane.  Such a brane configuration gives an $SO(8)^4$ enhanced gauge symmetry \Sen.  Now applying
a T-duality along the two directions of $T^2/\ZZ_2$ and then followed by an
S-duality, we obtain the heterotic theory on $K3 \times T^2$ with
$SO(8)^4$ gauge group \BD.

The duality can incorporate a non-zero four-form $G$-flux.  To
preserve supersymmetry, the $G$-flux in M-theory is required to be a
primitive $(2,2)$-form.  The $G$-flux is also quantized so that $G
\in H^{4}(Y,\ZZ)$.  Additionally, it must satisfy the constraint
(assuming no M2-branes) \refs{\BB, \SVW}
\eqn\Gcond{\frac{1}{2}\int_Y G\w G = \frac{\chi(Y)}{24}~.}

We shall take $G$ to be the exterior product of two (1,1)-forms,
one from each $K3$.  In the orbifold
limit, there are 19 primitive (1,1)-forms - 16 localized at each of
the fixed points and 3 nonlocalized ones.  They constitute the orthogonal Kummer basis, $\{\b_i,\g_I\}$ with $i=1,\ldots,16$,~$I=1,2,3$, and normalized to -2, for the primitive forms in $H^{(1,1)}(S)$.  Thus, the most general $G$-flux that we consider takes the form
\eqn\Gans{G=  C_{ij}\, \b_i \w \b'_j + C_{Ij}\, \g_I\w \b'_j +
D_{iJ}\,\b_i \w \g'_J + D_{IJ}\, \g_I \w \g'_J~,}
where we have placed primes to denote forms from the second $K3$ and $C_{ij}, C_{Ij}, D_{iJ}, D_{IJ}$ are integer constants suitably chosen such that $G$ is integral quantized and satisfies \Gcond.  The four different terms dualize to different types of fluxes in the heterotic theory.  Let us fix our convention by performing the duality operations always on the second $K3$.  Then, the $D_{IJ}$ and $D_{iJ}$ terms with nonlocalized (1,1)-forms $\g'_J$ dualize to give a non-zero torus curvature $\om$.  In contrast, $C_{ij}$ and $C_{Ij}$ terms with localized $\b'_j$ dualize to non-zero heterotic field strengths $F^j$.  The case of $D_{IJ}\neq 0$ in particular was discussed in detail in \refs{\dbeck,\BD,\beck,\BT}.

Of interest for us is the exchange of the two $K3$'s.  What we call the first or second $K3$ is certainly inconsequential for the physical theory.  But for the $G$-flux, such an exchange can interchange the different terms and result in a different heterotic dual theory when the ``second" $K3$ is dualized.  The set of $C_{ij}$ terms and also that of the $D_{IJ}$ terms map to itself under interchange.  However, for the other two sets of terms, we have $D_{iJ}\to C_{Ji}$ and $C_{Ij}\to D_{jI}$, which implies that two types of heterotic fluxes are exchanged under interchange of the two $K3$'s.  Hence, we shall focus below on fluxes of types $C_{Ij}$ and $D_{iJ}$.

For simplicity, let us take the $T^8$ covering space of $Y=T^4/\ZZ_2\times T^4/\ZZ_2$ to have standard periodicities $z_k \sim z_k + 1 \sim z_k + i$, for $k=1,\ldots,4$.  Let us begin first with the $D_{iJ}$ terms.  We write out explicitly the nonlocalized part
\eqn\GDij{G = D_{i1} \,\b_i \w \frac{1}{2}(\dzc\w \dbzd + \dbzc\w \dzd)
+ D_{i2}\, \b_i \w \frac{1}{2i}(\dzc\w \dbzd - \dbzc\w \dzd)~.}
Note that we have not included the term with $\g_3 = \frac{1}{2}(\dzc\w\dbzc - \dzd\w \dbzd)$ because it is not normalizable when the area of the torus fiber is taken to zero when mapping from M-theory to type IIB orientifold theory \refs{\dbeck,\BD}.
Now re-arranging the two terms, we have
\eqn\GDijj{\eqalign{G&= \frac{1}{2}(D_{i1} + iD_{i2}) \b_i\w \dbzc\w \dzd + \frac{1}{2}(D_{i1}-iD_{i2})\b_i\w\dzc\w\dbzd \cr
&=\frac{1}{2}[D_3\w\dzd +  \bD_3 \w \dbzd)\cr
&=\frac{1}{2}[D_3\w (dx_{10} + i dx_{11}) + \bD_3 \w (dx_{10} - i dx_{11})]\cr &=\frac{1}{2}(D_3 + \bD_3) \w dx_{10} + \frac{1}{2}(D_3 - \bD_3)\w i\,dx_{11}\cr
&=H_3\w dx_{10} + F_3\w dx_{11}~,} }
where we have introduced $D_3 = D_i \b_i \w \dbzc= (D_{i1}+ i D_{i2})\b_i \w \dbzc$ and substituted $z_4= x_{10} + i x_{11}$.   In particular,
\eqn\Hexp{H_3=dB_2=\frac{1}{2}(D_i\b_i \w \dbzc + \bD_i \b_i \w \dzc)=\frac{1}{2}d(\a \w \dbzc + \bal \w \dzc)} \eqn\Bexp{B_2=
\frac{1}{2}(\a\w \dbzc + \bal \w \dzc)~,}
where we have defined $D_i \b_i = d \a$.  Applying two T-dualities in the $z_3$ directions, the metric and the B-field of the type IIB theory get mixed.  After a further
S-duality, the resulting heterotic metric takes the form \refs{\dbeck,\BD,\BT}
\eqn\Jhet{ds^2 = e^{2\p}(\dza\dbza + \dzb\dbzb) + |\dzc + \a|^2~.}
This is the metric of the FSY solution in the $\ZZ_2$ orbifold limit. Thus, we see that
the M-theory solution, with a $G$-flux having a nonlocalized
(1,1)-form $\g'$ on the second $K3$ which we dualized, gets map to a heterotic
FSY solution.  Of course, one can check that the anomaly
condition \kfcon\ is satisfied.  This follows from \Gcond,
\eqn\Gconda{\eqalign{24=\frac{1}{2}\int_Y G\w G &=
\frac{1}{4}\,\int_S D_i\b_i\w\bD_j \b_j\int_S
\dzc\w\dbzc\w\dzd\w\dbzd\cr &=- \int_S D_i\b_i\w \bD_j\b_j~, }}
which is \kfcon\ after setting $\om = D_i\b_i= (D_{i1}+iD_{i2})\b_i$ and $F=0$.  Note that the torus bundle curvature here is localized at the fixed points of the base $T^4/\ZZ_2$.

Let us now return to the M-theory model and exchange the two $K3$'s.  The $G$-flux of \GDij\ maps to a $G$-flux with $C_{Ij}$ terms
\eqn\Gansb{\eqalign{G&= C_{1i}\frac{1}{2}(\dza\w \dbzb + \dbza\w\dzb) \w \b'_i
+ C_{2i}\frac{1}{2i}(\dza\w\dbzb - \dbzc\w\dzd)\w\b'_i\cr
&= [\frac{1}{2}(C_{1i}+iC_{2i})\dbza\w\dzb
+ \frac{1}{2}(C_{1i}-iC_{2i})\dza\w\dbzb]\w \b'_i\cr
&\equiv \frac{1}{2}[C_i \dbza\w \dzb + \bC_i \dza\w\dbzb] \w \b'_i\cr
&\equiv \frac{F^i}{4\pi} \w \b'_i
~. }}
Dualizing again in the $z_3, z_4$ directions, the resulting fluxes
on the type IIB orientifold are now very different.  With the $G$-flux localized at points on the second $K3$ as in \Gansb, we have in type IIB non-zero gauge field strengths $F^i \sim C_{Ii}\g^I$, on the D7/O7 planes \refs{\dbeck,\BD,\beck}.  Unaffected by the two
T-dualities and one S-duality, these gauge fluxes become the gauge field strengths of the
heterotic theory on $K3\times T^2$.  The anomaly condition again follows from \Gcond
\eqn\Gcondb{\eqalign{ 24=\frac{1}{2}\int_Y G\w G&= \frac{1}{32\pi^2}
\int_S F^i \w F^j \int_S\b'_i\w\b'_j\cr
&=  -\frac{1}{16\pi^2} \int_S F^i \w F^i~.}}

As an example, let us compare the models \fya\ and \kta\ presented in the last section.  In fact, it is easy to see that the FSY model of \fya\ arise from the $G$-flux
\eqn\tmodGa{G=(\b_1 + \b_2 - 2\b_3) \w \g'_1 +  (\b_1 + \b_2 - 2\b_3) \w \g'_2~,}
and that of \kta\ from
\eqn\tmodGb{G=(\g_1+\g_2)\w \b'_1 + (\g_1+\g_2)\w \b'_2 -2 (\g_1+\g_2)\w \b'_3~.}
They differ just by a switch of the two $K3$'s.

In short, we have shown that for identical $G$-flux \Gans\ and \Gansb\
which differs only in the assignment of what we called the first or second $K3$, the resulting dual heterotic models have very different background geometries.  The first maps to a FSY type solution and the second maps to a $U(1)$ gauge bundle on the K\"ahler manifold $K3\times T^2$.  Since the spectrum of the M-theory is invariant under the $K3$ exchange, it may seem that the distinct dual heterotic models must be physically identical.

This is however not the case in four dimensions.  We point out that such a heterotic-heterotic duality can only be apparent when both heterotic models are compactified further on an additional circle so that the external spacetime theory is three-dimensional.  The low energy theory would then be identical to that of M-theory on $K3\times K3$, which is also three-dimensional.  For in the process of ``dualizing" from M-theory to heterotic theory, we shrank down the area of the elliptic fiber, $A_{T^2}\to 0$, to obtain a type II and then later heterotic theory in four dimensions.  If we had not taken the zero area limit, then the exact duality would result in a seven-dimensional compact geometry, {\it e.g.} the FSY geomtry times an additional $S^1$.  More specifically, $A_{T^2} \sim 1/R_{S^1}$, and thus the $A_{T^2}\to 0$ limit corresponds to the decompactification limit of the extra circle in heterotic theory.

Taking into account of the deformation to
the orbifold limit discussed in Section~2, we have thus demonstrated a
connection between two smooth geometries, one K\"ahler and the other
non-K\"ahler, via a path through an orbifold limit on the moduli
space of the $K3$ surface.

\newsec{Discussions and outlook}

We have utilized the mapping to M-theory to connect K\"ahler and non-K\"ahler flux compactifications in heterotic theory.  The M-theory model organizes the heterotic models' torus and gauge bundle curvatures into a single four-form $G$-flux.  Turning on the torus and gauge bundle curvatures correspond to turning on different types of forms for the $G$-flux.  As we have shown, an exchange in the two $K3$'s with non-zero $G$-flux can lead to either the heterotic $K3\times T^2$ model or the FSY model.  Interestingly, M-theory on $K3\times K3$ can also be dual to Type IIA theory on $X_3 \times S^1$, where $X_3$ is a Calabi-Yau three-fold.  It has been pointed out in \AspinD\ that the exchange of two $K3$'s in this set up corresponds to mirror symmetry for the Calabi-Yau three-fold.  Following this observation, it is conceivable that the duality we have pointed to is related to a ``generalized" mirror symmetry in the heterotic theory with fluxes.

The transition we have discussed preserves $N=2$ supersymmetry in
four dimensions. An interesting old question is whether the space of all
$N=2$ string vacua is actually connected.  Prior to the current popularity of flux
compactifications, many works in the mid-90s gave evidences to unforseen connectedness between different vacua and thus hinted at such a possibility.  But with the recent increases in flux models which are quantized in integral units, hopes of a single connected moduli space have now faded.  But perhaps the simple well-known moduli space of $c=1$ closed bosonic string theory \refs{\DVV, \Ginsp} will give a guide to the moduli space of $N=2$ vacua.  In $c=1$, the moduli space includes the circle $S^1$ with radius $R_c$ and the
$\ZZ_2$ orbifold $S^1/\ZZ_2$ with radius $R_t$ for the target space
geometry.  Rather surprisingly, these two distinct geometries are
connected or dual at precisely the point $R_c/2= R_t = \sqrt{\a'}$.
In addition, there are three points in the moduli space which are
disconnected to all other theories.  Taking this example as a lead, we may
think that the moduli space of $N=2$ string vacua also smoothly
link together topologically distinct manifolds, but yet there will also
be regions of isolated vacua which do not have any moduli that
connect to the rest.  Our examples here of a connection between K\"ahler and non-K\"ahler geometries is an example of a somewhat surprising link.

Finally, we have for simplicity focused our attention on a subset of non-K\"ahler FSY solutions.  It would be interesting to explore the connectedness of the moduli space when the gauge bundle is non-Abelian.  Studying this may involve non-perturbative effects.  For instance on the M-theory side, we would need to consider M2-branes wrapping singular two-cycles of the $K3$ in order to generate non-Abelian flux.  On the heterotic side, wrapped branes of the sort discussed in \SY\ might also be required.  FSY solutions with torus curvature $\om$ having a (2,0) component should also be investigated for possible transitions.  These $N=1$ vacua would sit in the moduli space of all $N=1$ heterotic compactifications, which include the conventional Calabi-Yau compactifications.  This moduli space should incorporate the well-studied conifold transitions, both K\"ahler/K\"ahler and K\"ahler/non-K\"ahler types, which are $N=1$ transitions in heterotic theory.  We hope to explore some these issues in future works.

\bigskip\bigskip\bigskip

\centerline{\bf Acknowledgements}
\medskip
We would like to thank A.~Adams, C.~Beasley, K.~Becker, J.-X.~Fu,
J.~Lapan, B.~Lian, I.~Melnikov, E.~Sharpe, J.~Sparks, A.~Strominger and A.~Subotic for helpful discussions.  M.~Becker would like to thank the Harvard Department of Physics for warm hospitality and partial financial support while this work was carried out.  The work of M.~Becker is further supported in part by PHY-0505757, PHY-0555575, and the University of Texas A\&M.  The
work of of L.-S.~Tseng is supported in part by NSF grant DMS-0306600
and Harvard University. The work of S.-T.~Yau is supported in part
by NSF grants DMS-0306600, DMS-0354737, and DMS-0628341.

\bigskip\bigskip\bigskip

\appendix{A}{N=2 supersymmetry conditions}

We check that FSY geometry has $N=2$ supersymmetry in four
dimensions. Assuming $N=1$ SUSY, we derive the additional conditions
$N=2$ SUSY imposes.  Discussions and references on the $SU(2)$ structure relevant for
$N=2$ SUSY can be found in \refs{\GMW, \Dall}.  We start from the
supersymmetry constraints
\eqn\heta{\nabla_M \et + {1\o 8}\, H_{MNP}\,\g^{NP}\, \et=0~,}
\eqn\hetb{\g^M\pa_M\phi\, \et + {1\o 12}\,
H_{MNP}\,\g^{MNP}\,\et=0~,}
\eqn\hetc{\g^{MN}F_{MN} \et =0~, }
$N=1$ SUSY implies the existence of a no-where vanishing spinor
$\ea$ on the manifold $X$.  This gives an $SU(3)$ structure and in
particular we can define
\eqn\suth{J_{mn} = -i \ea^\dg \g_{mn} \ea~, \qquad\qquad
\O_{mnp} = e^{-2\p} \eab^\dg \g_{mnp} \ea~,} where the
complex conjugate spinor is defined to be ${\bar \et}_1 = B^*
\et_1^*$.\foot{The $B$ matrix here satisfies $B \g^m B^* = -
\g^*\,$ and we shall work in a  basis where $B^t = B$.}  The
manifold is required to be complex hermitian and the metric
conformally balanced.  Moreover, we have the relations
\eqn\noneeq{H=i(\bpa-\pa)J~, \qquad\qquad \O_{mnp}{\bar \O}^{mnp} =
8 e^{-4\p}~.}

$N=2$ SUSY implies the existence of a second no-where vanishing
spinor $\eb$. Both spinors have the same chirality, which we take to
be positive, and we shall assume that the two spinors are never
parallel.  We can therefore normalize so that
\eqn\snorm{\et^\dg_i \et_j = \d_{ij}~, \qquad i,j=1,2~.}
With the additional spinor,
there now exists a no-where vanishing one-form
\eqn\sone{v_m= \eab^\dg \g_m \eb~.} Alternatively, we can
write $\eb = \frac{1}{2} v_m \g^m \eab$.  In the holomorphic
coordinates $J_a{}^b = i \d_a{}^b$, we have $J_m{}^n v_n = i v_m\,$;
that is $v$ has only holomorphic components.  Moreover, the
normalization of $\eb$ implies that $|v|^2 = g^{a\bb}v_a\vb_\bb = 2$~.

We point out that the existence of a no-where vanishing one-form is
a strong constraint on $X$.  Specifically, from the Poincar\'e-Hopf
theorem, the number of zeroes of a vector field (and its dual
one-form) is at least that of the Euler characteristic ({\it i.e.}
$\geq|\chi|$).  Thus, $N=2$ SUSY having a non-vanishing one-form
requires $\chi(X)=0$.

Besides the vector, the additional spinor allows us to write down
the forms,
\eqn\twof{\eqalign{ \eb^\dg \g_{mn}\ea &= \frac{1}{2}
\vb^s\O_{smn}e^{2\p} \equiv  (K_2)_{mn} + i (K_1)_{mn}\cr \eb^\dg
\g_{mn} \eb  &= - i J_{mn} -2 v_{[m}\vb_{n]} \equiv -2i (K_3)_{mn} +
i J_{mn}}}
\eqn\threef{\eqalign{ \eab^\dg \g_{mnp} \ea & = (v \w
(K_2 + i K_1) )_{mnp} \cr \ebb^\dg \g_{mnp} \eb & = (v \w (K_2 - i
K_1) )_{mnp} \cr \ebb^\dg \g_{mnp} \ea & = (v \w J )_{mnp} \cr }}
Note that the hermitian form and holmorphic three-form can be
written as
\eqn\JO{\eqalign{ J &= K_3 + \frac{i}{2} v\w \vb \cr \O
&= e^{-2\p} (K_2 +i K_1) \w v~.}} The $K_A$'s, with $A=1,2,3\,$, reside in
a four-dimensional subspace and give a hyperk\"ahler structure
\eqn\sust{(K_A)_m{}^n (K_B)_n{}^k= -\d_{AB}[\d_m{}^k -
\frac{1}{2}(v_m\vb^k + \vb_m v^k)] + \e_{AB}{}^C (K_C)_m{}^k,} where
the additional terms in $v$'s are present since the forms are
defined in six-dimensions and not in four.  In particular, we have
for example $(K_1)_m{}^n(K_1)_n{}^ m = -4$  as typical for a
four-dimensional hyperk\"ahler space.

The first constraint equation \heta\ requires that all the
non-vanishing forms are covariantly constant with respect to the
$H$-connection.  Note that $N=1$ SUSY already ensures that $J$ and
$\O e^{2\p}$ are covariantly constantly. Hence, the additional
constraint of $N=2$ comes from requiring the covariantly constancy
of the one-form \eqn\vcovc{\na_m^H v_n = \na_m v_n -\frac{1}{2}
H_m{}^r{}_n v_r =0~.}

The second equation \hetb\ gives further differential constraints on
the forms. For $A=\g_{n_1\ldots n_p}$ antisymmetric combination of
gamma matrices, we can re-express \hetb\ as \GKMW
\eqn\hetbb{\pa_m\p\,
\et^\dg_j [A , \g^m]_\pm \et_i + \frac{1}{12} H_{mnp} \et^\dg_j [A,
\g^{mnp}]_\mp \et_i = 0} \eqn\hetbbb{\pa_m\p\, {\bar \et}^\dg_j [A ,
\g^m]_\pm \et_i + \frac{1}{12} H_{mnp} {\bar \et}^\dg_j [A,
\g^{mnp}]_\mp \et_i = 0}
where the $+$ or $-$ sign for the brackets
denotes symmetric or antisymmetric brackets, respectively. The
condition on the one-form $v$ can be derived with $A =  \g_{n_1n_2}$
resulting in the constraint
\eqn\onef{ d[e^{-2\p} v] + i * (H\w
e^{-2\p} v) =0~.}
For the two-forms and three-forms in \twof$\,$-\threef\ which we will
denote generically as $\c_2$ and $\c_3$, the conditions are
\eqn\twothree{\eqalign{ d[*(e^{-2\p} \c_2) ] & = 0 \cr d[*(e^{-2\p}
\c_3) ] & = 0}}

Now for the gauge field strength $F$, in addition to being
holomorphic $F^{(2,0)}=F^{(0,2)}=0$ and primitive $F^{mn}J_{mn}=0$,
we now have in general the condition, $F^{mn} (\c_2)_{mn} = 0$ which
gives the additional requirement
\eqn\gaugec{F^{mn} (K_3)_{mn} =0~.}

We now check that the FSY geometry with torus bundle curvature
$\om \in H^{(1,1)}(K3,\ZZ)$ satisfies the $N=2$ SUSY
conditions. First, note that as required, the Euler characteristic
of a $T^2$ bundle over $K3$ is zero. From the decomposition of \JO,
it is clear that we have the identification
\eqn\fyid{v =\th~,\qquad K_3= e^{2\p} J_{K3}~,\qquad K_2+iK_1 = e^{2\p}
\O_{K3}~.}
The $K_A$'s are thus the hyperk\"ahler forms on the
conformal $K3$.  We now check that the differential equations are
also satisfied.  First, the covariantly constancy as in \vcovc\ can
be shown using \noneeq\ and \JO\ to be equivalent to $\pa \th =0$.
Therefore, the torus bundle curvature twist can not contain a (2,0)
component.  The condition \onef\ can be shown to reduce to
\eqn\onefr{ \bpa_\bc \p\,\th^\bc = 0~, \qquad\qquad
H_{ab\bc}\,\th^\bc =0~.} This easily holds as $v^\bc= g^{\bc a}v_a=
(0,0,2)$.  The conditions of \twothree\ for various forms in \twof\
and \threef\ can also easily be checked to hold.  And lastly, for
the gauge field strength, we see that the additional requirement
\gaugec\ is satisfied since all field strengths are hermitian
Yang-Mills on the base $K3$.


\listrefs

\end